# SOUND ABSORPTION IN REPLICATED ALUMINUM FOAM

Arcady Finkelstein[1], Eugene Furman[1], Dmitry Husnullin[1], Konstantin Borodianskiy[2]

[1]Department of Foundry Engineering and Strengthening Technologies, the Ural Federal University, Yekaterinburg 620002, Russia
[2]Zimin Advanced Materials Laboratory, Department of Chemical Engineering, Biotechnology and Materials, Ariel University, Ariel 40700, Israel
* corresponding author: avinkel@mail.ru

## 1. Introduction

Sound absorption is an important technological task in machine-building and civil engineering. Porous materials are traditionally used for these purposes, as they are neither ignitable nor hygroscopic and thus suitable for noise oppression, first of all in means of transportation. Absorption of acoustic oscillation energy in porous metals occurs mainly due to viscous friction. A theoretical description of the process of energy viscous dissipation in a porous media on basis of Rayleigh classical model is given in paper [1], whereas the modern level of theory is set forth in Johnson-Champoux-Allard model [2]. Attempts of utilizing aluminum foam as the cheapest porous metal for sound absorption are related to forming of the open porous structure by rolling [3] or by heat treatment [4]. However, the sound absorption ratio of metal foam presented in these papers does not rise over 80%, whereas it reaches 99.9% in a wide frequency range when we take conventional sound-absorption materials (i.e. glass-wool). The problem of foamed metal consists of considerable reflection of acoustic waves from the surface.

Replicated aluminum foam [5] (fig.1) does not possess closed porosity at all and has much larger gap, which provides the level of reflection lower than that of metal foams, so replicated aluminum foam is a promising sound-absorption material. Fernández et.al studied dependency of sound absorption coefficient at different frequencies on the bed fraction, plate thickness and depth of air gap of replicated aluminum foam produced by infiltration of pre-sintered filling material under high pressure [6]. Han et.al found dependency of sound absorption ratio on the bed fraction and plate thickness of replicated aluminum foam within the technology of vacuum infiltration of pre-sintered bed [7]. The sound absorption coefficient of replicated aluminum foam reached 98%. The technology of loose bed vacuum infiltration allows managing geometric parameters of the porous structure [8]. Aluminum is promised material in today's metallurgy industry [9-12] so it becomes possible to predict the acoustic behavior of replicated aluminum foam which typical structure shown in Fig. 1.

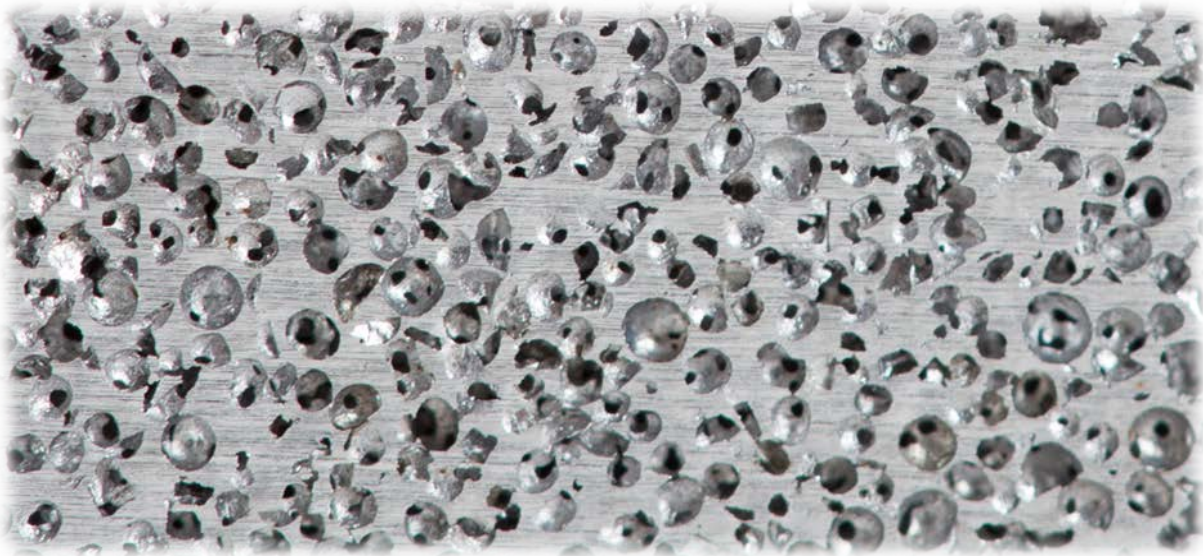
**Fig. 1.** Replicated aluminum foam structure. Magnification X10.

## 2. Experimental

*2.1. Theoretical studies*

The classical model has been taken from the work of Cremer [1] as a basis of theoretical calculation of sound absorption in replicated aluminum foam, while geometrical parameters of a porous structure conform to the previously obtained model in [8]. To provide a picture of propagation of acoustic waves of low and medium intensity (acoustic Mach number $M_{ac}<0.5$) in the air, sound absorption effects can be well described by the following wave equation:

$$\rho \frac{\partial^2 p}{\partial t^2} = K \frac{\partial^2 p}{\partial x^2} \qquad (1)$$

$$K = \rho \frac{\partial p}{\partial \rho} \qquad (2)$$

where ρ is the media density, K is bulk modulus of the media elasticity, and p is an instant value of sound pressure. The equation has an analytical solution conforming to propagation of decaying plane sine waves in the media; the analysis of this equation shows that the relation of sound pressure to oscillating particle velocity is a constant value and is referred as wave impedance of the media being one of wave parameters of the media:

$$W = \frac{p}{v} \qquad (3)$$

The second wave parameter of the media is the propagation constant which characterizes sound wave decay in the media:

$$\gamma = \beta + ik \qquad (4)$$

where $\beta$ is the rate of decay, $k$ is the acoustic wave number, and $i$ is the imaginary unit $\sqrt{i^2} = -1$.

Sound absorption properties of a material are characterized by the sound absorption coefficient:

$$\alpha = \frac{I_{fall} - I_{ref}}{I_{fall}} \tag{5}$$

where $I_{fall}$ is the incident wave intensity; $I_{ref}$ is the reflected wave intensity. The value shows a proportion of sound energy absorbed by the material or structure. The sound absorption coefficient can also be expressed via dimensionless impedance of the material or structure:

$$\alpha = 1 - \left|\frac{\bar{Z}-1}{\bar{Z}-1}\right|^2 \tag{6}$$

$$\bar{Z} = \frac{Z}{W_0} \tag{7}$$

where $\bar{Z}$ is dimensionless impedance of the material or structure (from here on: values with upper streak are neon-dimensional zed parameters in relation to air properties), $Z$ is impedance of the material or structure, $W_0$ is air wave impedance. $\bar{Z} = \bar{X} + i\bar{Y}$ is a complex value.

Equation (6) canbere-written with the use of present values:

$$\alpha = \frac{4\bar{X}}{(\bar{X}+1)^2 + \bar{Y}^2} \tag{8}$$

Two variants of sound absorption structures have been studied in this survey: a porous material on a rigid wall and a porous material with air gap between one and rigid wall (Fig. 2).

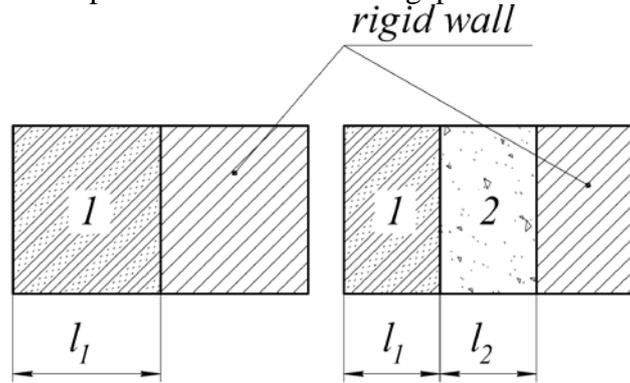

**Fig. 2.** Patterns of sound absorption structures for calculation of dimensionless impedance: 1: replicated aluminum foam, 2: air.

In the first example, the present and imaginary parts of the dimensionless impedance should be calculated according to the following equations:

$$\bar{X} = \frac{\bar{W}_x \sinh(2\beta l_1) + \bar{W}_y \sin(2kl_1)}{\cosh(2\beta l_1) - \cos(2kl_1)} \tag{9}$$

$$\bar{Y} = \frac{\bar{W}_y \sinh(2\beta l_1) - \bar{W}_x \sin(2kl_1)}{\cosh(2\beta l_1) - \cos(2kl_1)} \tag{10}$$

In the second example, there land imaginary parts of the dimensionless impedances should be calculated according to the following equations:

$$2\bar{X} = \frac{\begin{array}{c}\overline{W}_x(a^2 + |\overline{W}|^2)\sinh(2\beta l_1) + \overline{W}_y(a^2 - |\overline{W}|^2)\sin(2kl_1) - \\ -2a\overline{W}_x\overline{W}_y(ch(2\beta l_1) - \cos(2kl_1))\end{array}}{\begin{array}{c}a^2(ch(2\beta l_1) - \cos(2kl_1)) + 2a(\overline{W}_x \sin(2kl_1) \\ -\overline{W}_y \sinh(2\beta l_1)) + |\overline{W}|^2(\cosh(2\beta l_1) + \cos(2kl_1))\end{array}} \quad (11)$$

$$2\bar{Y} = \frac{\begin{array}{c}\overline{W}_x(|\overline{W}|^2 - a^2)\sin(2kl_1) + \overline{W}_y(|\overline{W}|^2 + a^2)\sinh(2\beta l_1) - \\ -2a(\overline{W}_x^2 \cos(2kl_1) + \overline{W}_y^2 \cosh(2\beta l_1))\end{array}}{\begin{array}{c}a^2(ch(2\beta l_1) - \cos(2kl_1)) + 2a(\overline{W}_x \sin(2kl_1) - \\ -\overline{W}_y \sinh(2\beta l_1)) + |\overline{W}|^2(\cosh(2\beta l_1) + \cos(2kl_1))\end{array}} \quad (12)$$

where $a = \cot(k_0 l_2)$; $k_0$ is the wave number in air; $l_1$ and $l_2$ – see Fig. 1; $\overline{W}_x$, $\overline{W}_y$ are real and imaginary parts of wave impedance of replicated aluminum foam; $|\overline{W}|$ is the modulus of wave impedance of replicated aluminum foam. Thus, the task boils down to calculation of real and imaginary parts of complex values W and γ for replicated aluminum foam as per dependencies [1]:

$$\overline{W}_x = \frac{1}{\sigma}\sqrt{\frac{\chi æ}{2}(1 + \sqrt{1 + \eta^2})} \quad (13)$$

$$\overline{W}_y = -\frac{\eta}{\sigma}\sqrt{\frac{\chi æ}{2}\frac{1}{(1 + \sqrt{1 + \eta^2})}} \quad (14)$$

$$\bar{\beta} = \eta\sqrt{\frac{\chi}{2æ}\frac{1}{(1 + \sqrt{1 + \eta^2})}} \quad (15)$$

$$\bar{k} = \sqrt{\frac{\chi}{2æ}(1 + \sqrt{1 + \eta^2})} \quad (16)$$

where æ is the elasticity ratio is taken as equal to 1 as per [1]; $\chi$ is the structure factor. Air compression in pores is characterized with relation of effective density to free air density (structure factor $\chi = \rho_{эф}/\rho_o$). It is obvious that the ratio of maximum to minimum pore sizes has impact on air compression. As per [8], the structure factor for replicated aluminum lies within values 2-9 and is determined by the bed fraction, porosity and infiltration pressure.

$\sigma$ is the ratio of the area of transparent pores exposed on the sample surface and the overall area of the outer side of the porous sample; it is determined by the structured sphere model (fictitious ground) of packed bed and depends on porosity. For replicated aluminum foam it varies in range 0.70-0.79 [13].

$\eta$ is the parameter characterizing ration between friction resistance and inertial resistance of air in pores; it is determined by the following equation:

$$\eta = \frac{r\sigma}{2\pi f \chi \rho_0} \quad (17)$$

where $f$ is oscillation frequency; $r$ is flow resistivity. The latter should be defined as:

$$r = \frac{\mu}{K_{perm}} \quad (18)$$

where $\mu$ is dynamical viscosity; $K_{perm}$ is replicated aluminum foam permeability. The latter is determined basing on parameters of the porous structure by the theoretical dependence.

A theoretical calculation of the sound absorption factor was realized in Microsoft Excel spreadsheet processor by the above mentioned scheme.

*2.2. Sample preparation and experimental*

Samples were prepared by the loose NaCl bed vacuum infiltration technology. Samples were made of Al alloy A356, 80 mm dia. in accordance with the factor sunder study:
1. Fraction of the bed used: 0.4-0.63 mm, 0.8-1 mm, 1.5-2 mm.
2. Thickness of samples: 10 and 20 mm.
3. Porosity: 53-57% (basic, loose bed), 58-62% (bed compaction by vibration), and 66-70% (compaction by vibration with addition of 0.2% graphite powder). Porosity of samples was measured indirectly by weighing (specific weight of the initial aluminum alloy was determined in advanced).
4. Air pressure drop at infiltration defining permeability factor: 1 bar and 0.25 bar.

Sound absorption coefficient at normal incidence was measured by means of a sound level meter system made by Bruel & Kjaer Company based on the acoustic spectrum analyzer 2144 in the research center of AvtoVAZ JSC (Samara, Russia) in an impedance tube similarly to [6], with an air gap of 20 mm deep and without it. A mean arithmetic value of three experimental samples is related to each point.

**3. Results and Discussion**

The conducted experiments (Fig. 3-7) showed, similarly to the data of [7], a very high sound absorption ratio, up to 99% in a narrow frequency range. If an air gap is presents due to the interference effects, sound absorption will increase in a wide frequency range, whereas the position of frequency maximums will shift toward low frequencies and an additional maximum will be generated at high frequency which is well confirmed by the Cremer model [1]. The highest sound absorption factor is achieved at the minimum interrelation of maximum and minimum pore sizes, which is confirmed by increase of sound absorption of replicated aluminum foam both as the bed fraction decreases and as the pressure drop decreases. Based on this conclusion, one can suppose that the low sound absorption ratio in [6] is a consequence of the large pressure drop as infiltration is performed (the value of this pressure drop is not identified in the presented work), and thus, a consequence of the large interrelation of maximum and minimum pore sizes. Increase of a sample thickness will shift the maximum of sound absorption factor towards low frequencies, whereas its absolute value will decrease. Impact of porosity is also obvious, where of a gap value will increase, i.e. reflection of acoustic waves from a surface of a porous plate will decrease which causes the sound absorption factor to increase. Porosity increase will also slightly shift the maximum of sound absorption toward high frequencies.

a)

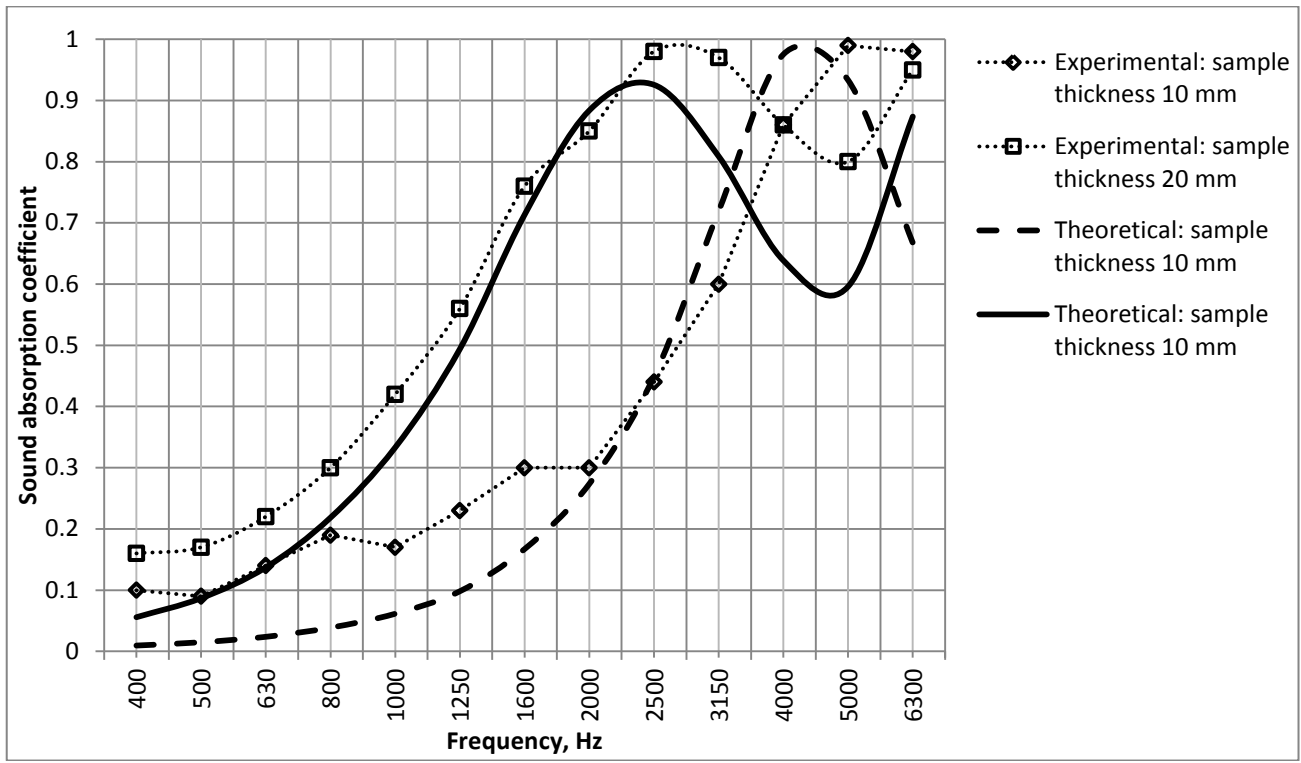
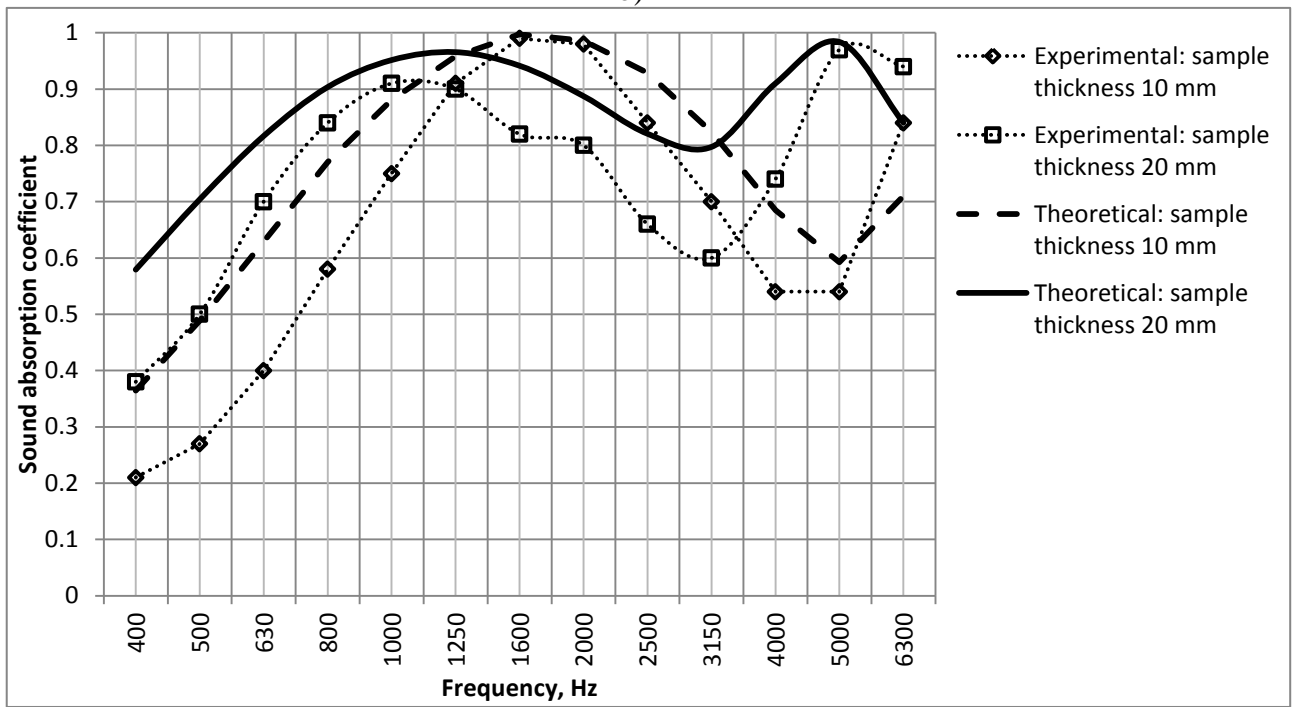

**Fig. 3.** The samples are made of replicated aluminum foam with 0.4-0.63 mm pore size and 60% porosity under 0.5 bar infiltration pressure drop. (**a**) Sound absorption coefficient dependency on frequency of samples with different thickness without air gap; (**b**) Sound absorption coefficient dependency on frequency of samples with different thickness with 20 mm air gap.

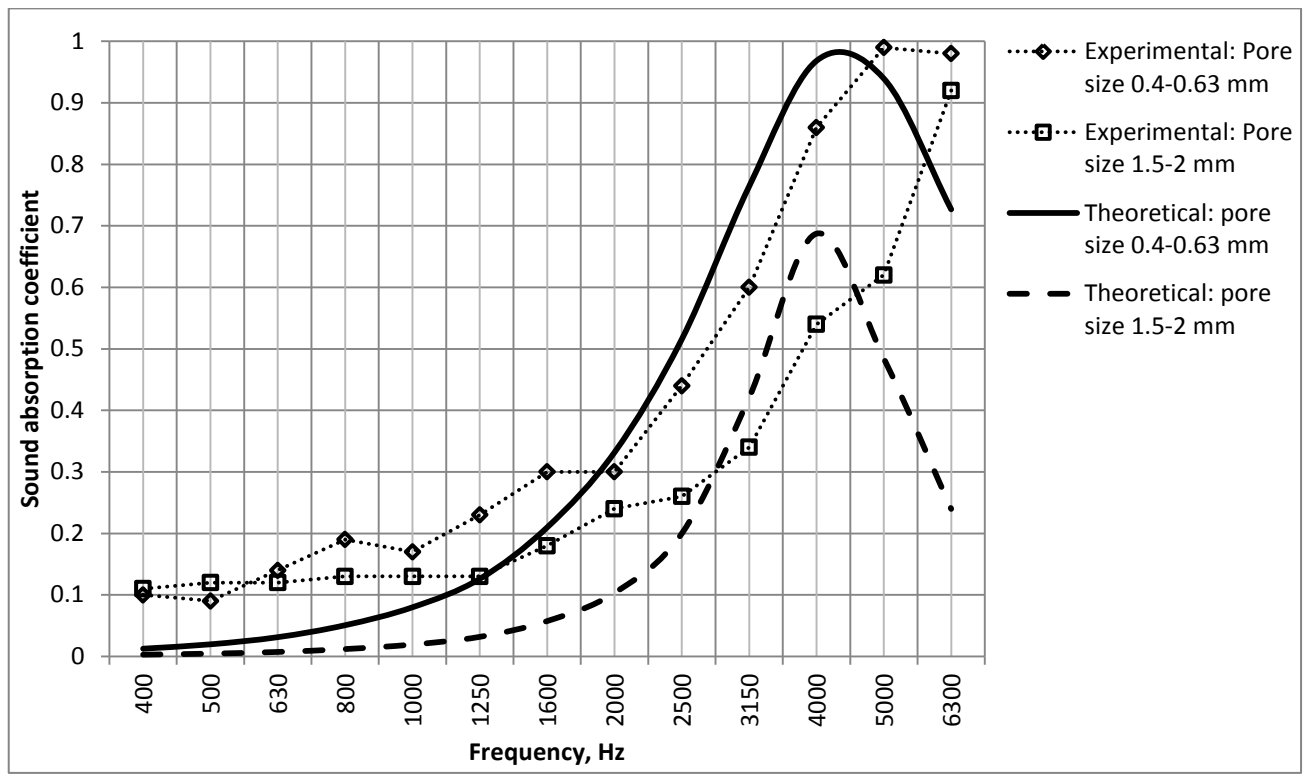

b)

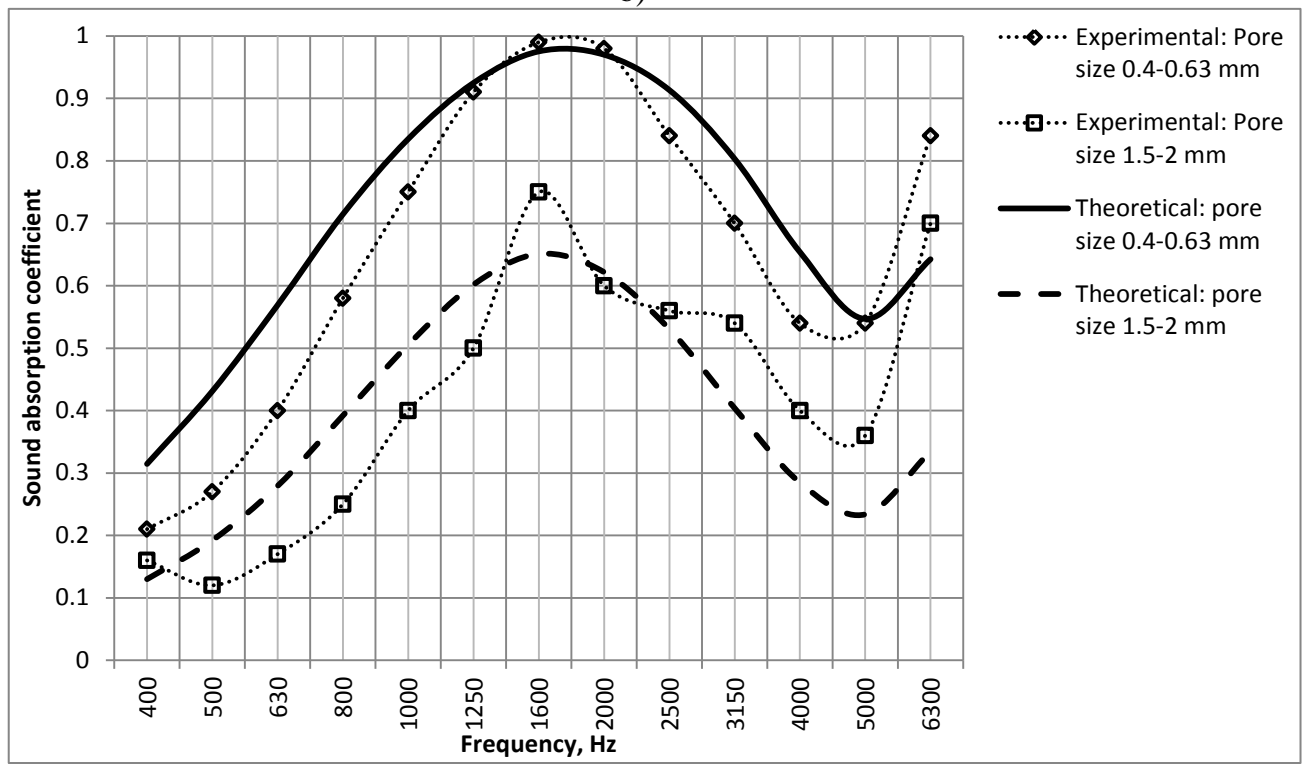

**Fig. 4.** The samples are made of replicated aluminum foam with 60% porosity and 10 mm thickness under 0.5 bar infiltration pressure drop. (**a**) Sound absorption coefficient dependency on frequency of samples with different mean pore diameter without air gap; (**b**) Sound absorption coefficient dependency on frequency of samples with different mean pore diameter with 20 mm air gap.

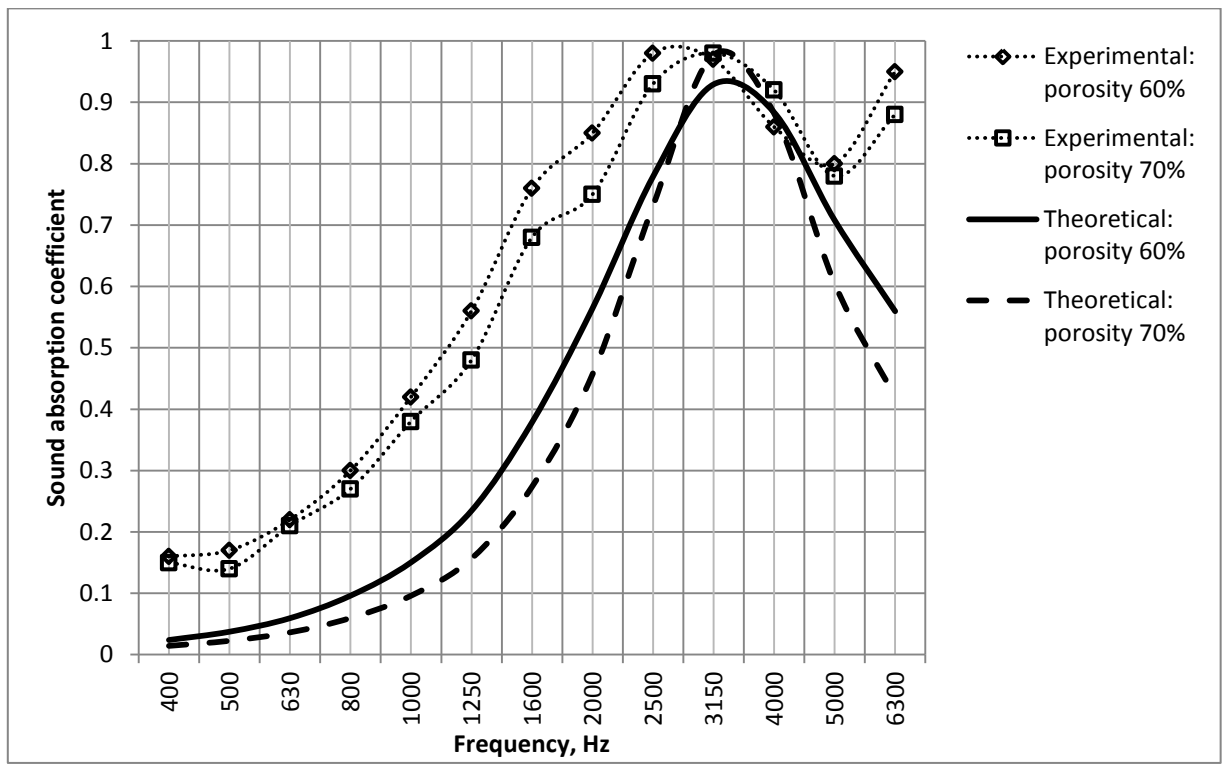

b)

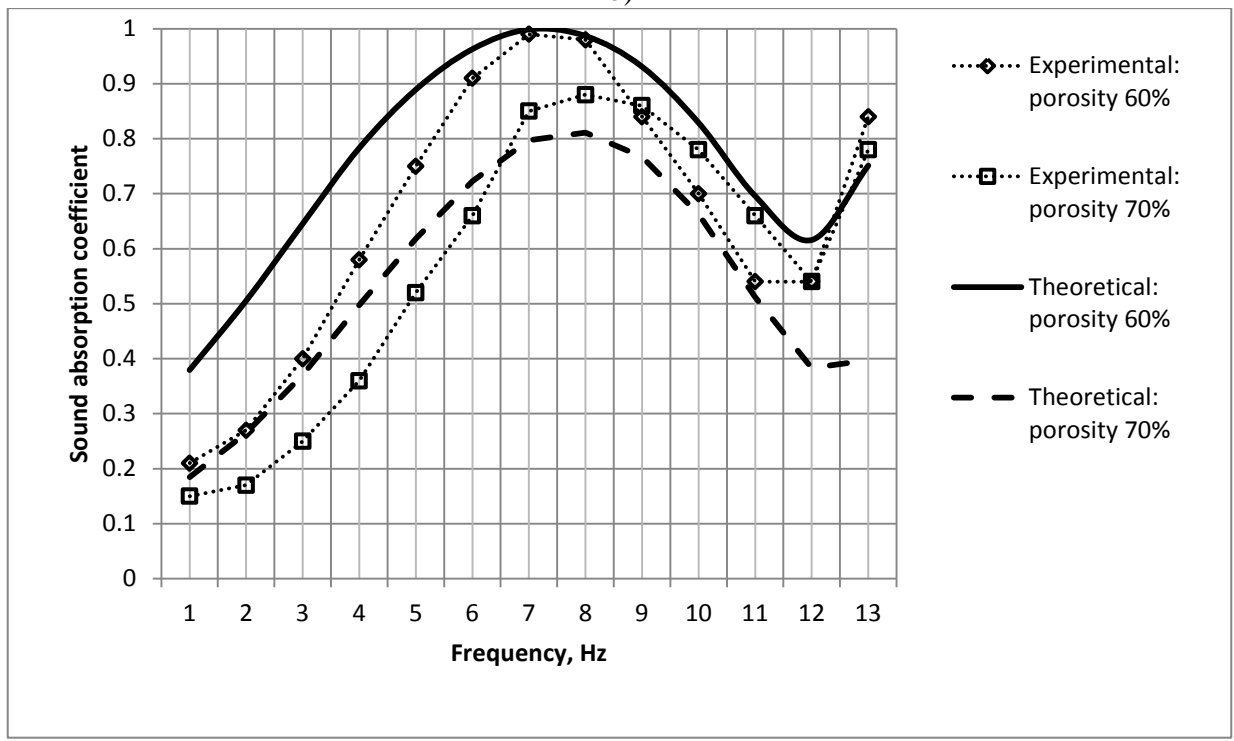

**Fig. 5.** The samples are made of replicated aluminum foam with 0.4-0.63 mm pore size and 10 mm thickness under 0.5 bar infiltration pressure drop. (**a**) Sound absorption coefficient dependency on frequency of samples with different porosity without air gap; (**b**) Sound absorption coefficient dependency on frequency of samples with different porosity with 20 mm air gap.

1-75

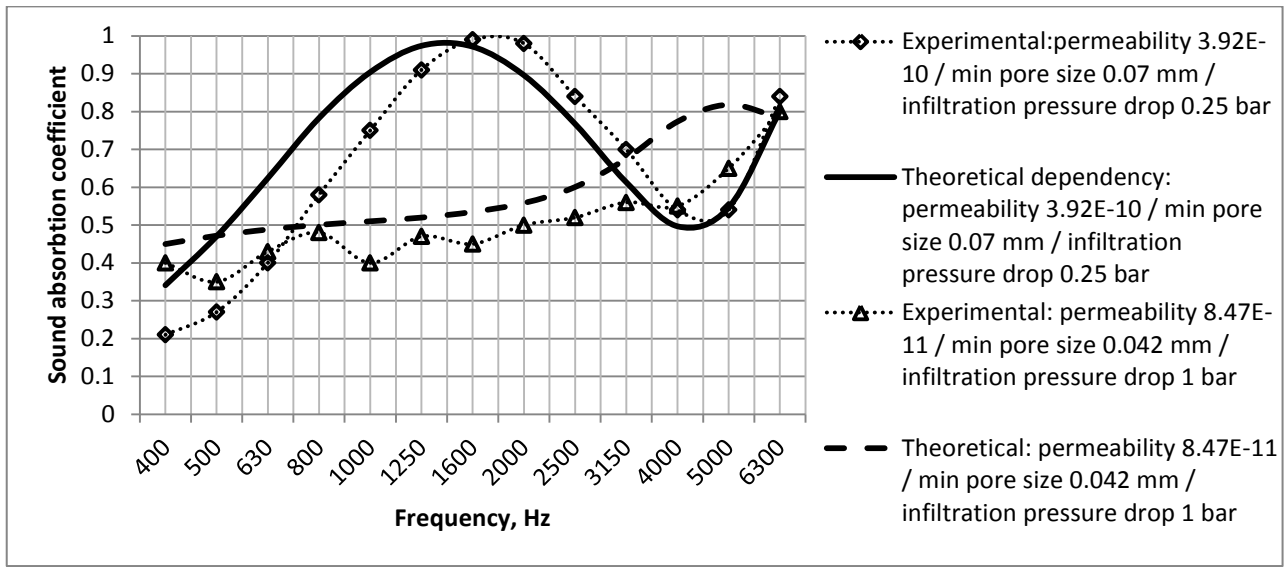

**Fig. 6.** Theoretical and experimental sound absorption coefficient dependency on frequency for samples with different permeability. The samples are made of replicated aluminum foam with 60% porosity, 0.4-0.63 mm pore size and 10 mm thickness. Samples tasted on 20 mm air gap apparatus.

Using the data presented in [6, 7] concerning sound absorption for the theoretical analysis will not seem possible, as these works do not identify such important structural parameters of a porous media as the pore minimum size and the permeability factor which depends on it. The porous structure model [8] used for a calculation pattern taking into account only air viscous resistance to friction is well confirmed by experimental data. The calculated sound absorption value occurs to be somewhat lower than the experimental one, probably due to energy heat dissipation in a porous media, as it is shown in [2]. However, Cremer's model [1] well predicts a frequency position of sound absorption maximum, while using of the model [2] requires determining not only geometric but also heat-geometric parameters of a porous structure, which does not seem rational, as it makes the model more complicated and does not provide any practical effect. According to the experience-acknowledged model, the maximum sound absorption will be provided by using of a fine bed fraction with the minimum pressure drop and maximum bed porosity. To regulate the frequency maximum of sound absorption, it is rational to use the porous layer thickness (Fig. 7) and the depth of air gap (Fig. 8).

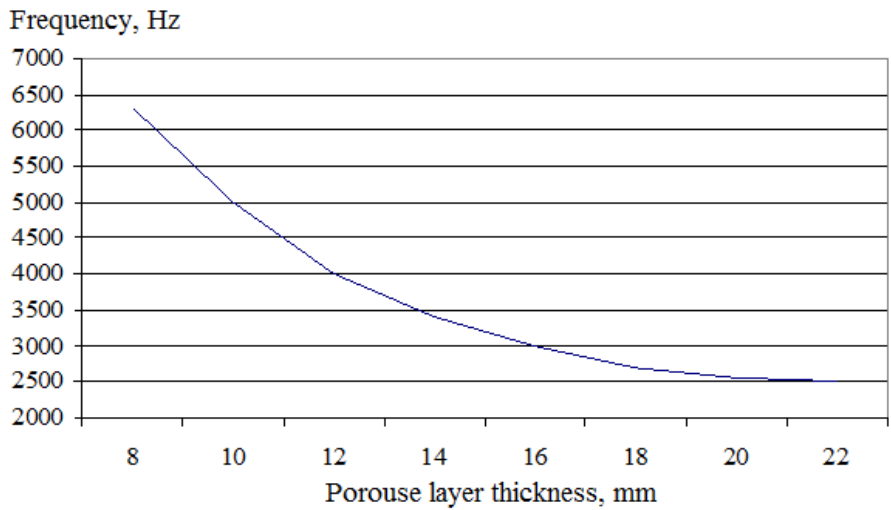

**Fig. 7.** Influence of porous layer thickness on a sound absorption maximum.

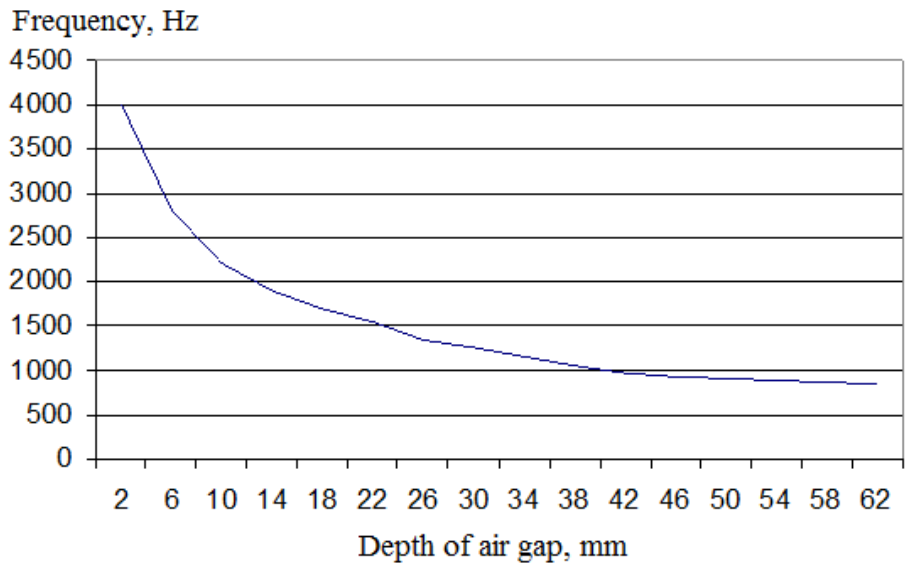

**Fig. 8.** Influence of depth of air gap on sound absorption maximum.

**4. Conclusions**

The derived dependencies are used for designing of sound-absorptive plates made of replicated aluminum foam at Composite Materials LLC (Kirovgrad, Russia) for industrial equipment emitting noise in a narrow frequency range.

**Acknowledgments**

The article is composed basing on results of the researches conducted within realization and at the expense of funding of Improving Competitiveness Program "5-100-2020". The work was supported by Act 211 Government of the Russian Federation, contract № 02.A03.21.0006.

**References**

1. L. Cremer, Die wissenschaftlichen Grundlagen der Raumakustik. 3rd ed. Hirzel Verlag: Leipzig, **1950**.

2. Y. Champoux, J.F. Allard, Dynamic tortuosity and bulk modulus in air-saturated porous media. *J Appl Phys* **1991**, 70 (4), 1975–1979.
3. T.J. Lu, A. Hess, M.F. Ashby, Sound absorption in metallic foams. *J Appl Phys* **1999,** 85 (11), 7528–7539.
4. A. Byakova, Y. Bezim'yanny, S. Gniloskurenko, T. Nakamura, Fabrication method for closed-cell aluminum foam with improved sound absorption ability. *Procedia Mater Science* **2014**, 4, 9–14.
5. H.A. Kuchek, Method of producing clad porous metal articles. Patent US 3138856, **1964**.
6. P.M. Fernández, L.J. Cruz, L.E. García Cambronero, C. Díaz, M.A. Navacerrada, Sound absorption properties of aluminum sponges manufactured by infiltration process. *Adv Mater Res* **2011**, 146, 1651–1654.
7. F. Han, G. Seiffert, Y. Zhao, B. Gibbs, Acoustic absorption behavior of an open-celled aluminum foam. *J Appl Phys D* **2003,** 36 (3), 294–302.
8. E. Furman, A. Finkelstein, M. Cherny, Permeability of aluminum foams produced by replication casting. *Metals* **2013**, 3 (1), 49–57.
9. V. Selivorstov, Y. Dotsenko, K. Borodianskiy, Influence of Low-Frequency Vibration and Modification on Solidification and Mechanical Properties of Al-Si Casting Alloy, *Materials* **2017**, 10 (5), 560-570.
10. A. Finkelstein, A. Schaefer, O. Chikova, K. Borodianskiy, Study of Al-Si Alloy Oxygen Saturation on its Microstructure and Mechanical Properties, *Materials* **2017**, 10 (7), 786-794.
11. K. Borodianskiy, V. Selivorstov, Y. Dotsenko, M. Zinigrad, Effect of additions of ceramic nanoparticles and gas-dynamic treatment on Al casting alloys, *Metals* **2015**, 5, 2277-2288.
12. V. Selivorstov, Y. Dotsenko, K. Borodianskiy, Gas-dynamic influence on the structure of cast of A356 alloy, *Herald of the Donbass State Engineering Academy*. Collection of science papers **2010**, 3 (20), 234-238.
13. A. Finkelstein, Replicated aluminum foam, theory and practice: Doctoral Thesis /Russia: Ural State Technical University, **2010**.